\documentclass[english,aps,manuscript]{revtex4}
\usepackage[T1]{fontenc}
\usepackage[latin9]{inputenc}
\usepackage[a4paper]{geometry}
\usepackage{amsmath}
\usepackage{amssymb}
\usepackage{graphicx}

\makeatletter

\newcommand{\lyxdot}{.}

\@ifundefined{textcolor}{}
{%
 \definecolor{BLACK}{gray}{0}
 \definecolor{WHITE}{gray}{1}
 \definecolor{RED}{rgb}{1,0,0}
 \definecolor{GREEN}{rgb}{0,1,0}
 \definecolor{BLUE}{rgb}{0,0,1}
 \definecolor{CYAN}{cmyk}{1,0,0,0}
 \definecolor{MAGENTA}{cmyk}{0,1,0,0}
 \definecolor{YELLOW}{cmyk}{0,0,1,0}
 }

\makeatother

\usepackage{babel}
\begin{document}

\title{A model for red blood cells in simulations of large-scale blood flows}

\author{Simone Melchionna}

\affiliation{CNR-IPCF, Consiglio Nazionale delle Ricerche, P.le A. Moro 2, 00185,
Rome, Italy}

\date{28.1.2011}
\begin{abstract}
Red blood cells (RBCs) are an essential component of blood. A method
to include the particulate nature of blood is introduced here with
the goal of studying circulation in large-scale realistic vessels.
The method uses a combination of the Lattice Boltzmann method (LBM)
to account for the plasma motion, and a modified Molecular Dynamics
scheme for the cellular motion. Numerical results illustrate the quality
of the model in reproducing known rheological properties of blood
as much as revealing the effect of RBC structuring on the wall shear
stress, with consequences on the development of cardiovascular diseases.
\end{abstract}
\maketitle

\section{Introduction}

The study of blood flows in physiological vessels, named hemodynamics,
is an active field of research from a fundamental point of view and
for the strategic medical implications related to cardiovascular diseases.
Atherosclerosis, for instance, is the leading cause of death in western
countries and its societal impact is constantly increasing \cite{_heart_2009}.
Atherogenesis is the formation of plaques within the endothelium,
the tissue forming the arterial walls, and is triggered by low levels
of shear stress at the wall region. The biomechanical origins of atherogenesis
relate to the disturbed flow patterns encountered in close proximity
of the endothelium \cite{rybicki_prediction_2009}.

Hemodynamics entails several physical and biochemical levels and understanding
blood flows in complex geometries, from the large-scale coronary arteries
to microcapillaries, has attracted considerable attention over centuries,
starting from the early investigations by Leonardo da Vinci \cite{skalak_asme_1981}.
Blood exhibits both visco-elastic and thyxotropic response and, most
importantly, shear-thinning, occurring at shear rates commonly encountered
in flow conditions corresponding to large-scale arteries, such as
in the coronary system. 

In the last decades, computer simulation has provided new understanding
on the flow patterns of blood modeled as a continuum. In physiological
condition, however, blood presents a large volume fraction of red
blood cells (RBC), with hematocrit level $H$ ranging between $35$
and $50\,\%$. Recent computational models have been put forward to
reproduce blood from a bottom-up perspective, that is, by including
plasma and individual red blood cells, either as flexible bodies \cite{bagchi_mesoscale_2007,noguchi_shape_2005,macmeccan_simulating_2009,sun_particulate_2005,fedosov_multiscale_2010,zhang_immersed_2007,dupin_modeling_2007}
or as rigid ellipsoids \cite{janoschek_simplified_2010}. A detailed
representation of RBCs is crucial to study microcirculation in capillaries,
a situation where shape, deformability and near-field hydrodynamic
response of single red blood cells need to be accurately accounted
for. Conversely, not much effort has been devoted to modeling red
blood cells in situations where the far-field hydrodynamics and the
global rheological response of blood are taken into account, two aspects
that can have strong impact on large-scale flows. Such possibility
would enable to study blood flows in large cardiovascular networks,
entailing from the red blood cell level up to heart-size geometries,
with a computational effort that represents a trade-off between physical
fidelity and computational feasibility.

In this paper we address the issue of designing a robust physical
representation of blood by a model that can be used in large-scale
coronary arteries. Therefore, our model accounts for the role played
by RBCs in the non-Newtonian hemorheology together with accounting
for cell crowding in proximity of the vessel walls. This aspect is
expected to have dramatic consequences on the distribution of endothelial
shear stress, due to the inhomogeneous distribution of red blood cells
in proximity of the arterial wall.

Red blood cells are globules that present a biconcave discoidal form,
and a soft membrane that encloses a high-viscosity liquid made of
hemoglobin. In presence of an external shear field, red blood cells
present both a solid-like tumbling and a vesicular motion with the
attendant sliding of the membrane, the so-called tank treading motion.
As a result, RBC encodes both rotational and orientational responses
that deeply and non-trivially modulate blood rheology. We account
for these contributions and neglect memory-related response of the
cell membrane arising from the rearrangement of the cytoskeleton.
In addition, we take into account some degree of cell softness for
the RBC-RBC collisional properties and disregard global shape rearrangements. 

The proposed model focuses on three independent components: the far-field
hydrodynamic interaction of a RBC in a plasma solvent, the raise of
viscosity of the suspension with the hematocrit level and the many-body
collisional contributions to viscosity. These three critical ingredients
conspire to produce large-scale hemorheology and the local structuring
of RBCs. 

The paper is organized as follows. In Section \ref{sec:Computational-Model}
the computational model is described by focusing on blood plasma at
first, and then incrementally introducing the RBC hydrodynamic interactions
(roto-translational and tank treading coupling), the modulation of
viscosity of the suspension and the RBC-RBC mechanical forces. Section
\ref{sec:Numerical-Results} illustrates the numerical validation
of the model and presents results on RBC-rich blood flows in a realistic
vessel.

\section{Computational Model\label{sec:Computational-Model}}

\subsection{Blood plasma}

More than $99\,\%$ in volume of blood is composed by plasma and red
blood cells, where the latter can account up to $50\,\%$ in volume
of the suspension. Plasma is a water-like Newtonian fluid that can
be modeled as a continuum, with the Navier-Stokes equations governing
the macrodynamics. 

For the purpose of reproducing the plasma evolution in high complex
geometries (and vanishing hematocrit) together with the inclusion
of red blood cells as suspended bodies, a convenient route is to model
plasma at kinetic level via the Lattice Boltzmann (LB) method \cite{succi_lattice_2001}.
The LB method deals with the evolution of the distribution function
$f({\bf x},{\bf v},t)$, describing the probability to find a plasma
particle at site ${\bf x}$, velocity ${\bf v}$ and time $t$, and
recovers Newtonian rheology in the limiting macroscopic space/time
scale. By representing the distribution function in discretized form
over a cartesian mesh ${\bf x}$, $f({\bf x},{\bf v},t)\rightarrow f_{p}({\bf x},t)$,
where the subscript $p$ labels a set of discrete speeds ${\bf c}_{p}$
connecting mesh points to mesh neighbors, the evolution dynamics over
a timestep $h$ reads
\begin{eqnarray}
f_{p}({\bf x}+h{\bf c}_{p},t+h) & = & f_{p}^{*}({\bf x},t)\label{eq:streaming}
\end{eqnarray}
with $f_{p}^{*}({\bf x},t)$ being the post-collisional population,
\begin{eqnarray}
f_{p}^{*} & = & (1-\frac{h}{{\cal T}})f_{p}+\frac{h}{{\cal T}}f_{p}^{eq}+h\Delta f_{p}^{drag}\label{lbe_discrete}
\end{eqnarray}

${\cal T}$ is a characteristic relaxation time and $f_{p}^{eq}$
the Maxwellian equilibrium expressed as a second-order low-Mach expansion
in the fluid velocity ${\bf u}$, 
\begin{equation}
f_{p}^{eq}=w_{p}\rho\left[1+\frac{{\bf u}\cdot{\bf c}_{p}}{c^{2}}+\frac{({\bf u}\cdot{\bf c}_{p})^{2}-c^{2}u^{2}}{2c^{4}}\right]
\end{equation}

Eqs.~(\ref{eq:streaming}-\ref{lbe_discrete}) encode the effect
of streaming, that is, the kinematic motion of free particles along
straight trajectories, together with the solvent-solvent and the solvent-solute
{}``molecular'' collisions. The solvent-solvent collisions are included
via the so-called BGK relaxation kernel, expressed by the fact that
the distribution strives to reach the local equilibrium over the timescale
${\cal T}$. The plasma kinematic viscosity $\nu$ relates to ${\cal T}$
via 
\[
\nu=c^{2}({\cal T}-h/2)
\]
where $c$ is the plasma sound speed.

In this work, we employ the D3Q19 lattice scheme, where one has $c=1/\sqrt{3}$
, and $w_{p}$ stands for a set of normalized weights with $p=0,...,18$,
being equal to $w_{p}=1/3$ for the population corresponding to the
null discrete speed ${\bf c}_{0}=(0,0,0)$, $w_{p}=1/18$ for the
ones connecting first mesh neighbors ${\bf c}_{1,...,6}=(\pm1,0,0),\,(0,\pm1,0),\,(0,0,\pm1)$,
and $w_{p}=1/36$ for second mesh neighbors, ${\bf c}_{7,...,18}=(\pm1,\pm1,0),\,(\pm1,0,\pm1),\,(0,\pm1,\pm1)$. 

The term $\Delta f_{p}^{drag}$ accounts for the presence of suspended
RBC that act as body forces on the plasma in a hydrokinetic way, as
detailed out in the next section. It has the following expression
\begin{equation}
\Delta f_{p}^{drag}=hw_{p}\rho\left[\frac{{\bf G}\cdot{\bf c}_{p}}{c^{2}}+\frac{({\bf G}\cdot{\bf c}_{p})({\bf u}\cdot{\bf c}_{p})-c^{2}{\bf G}\cdot{\bf u}}{2c^{4}}\right]\label{eq:chen_force}
\end{equation}
where ${\bf G}$ is the local body-fluid coupling. Eq. (\ref{eq:chen_force})
is a representation of the body force consistent with the second-order
Hermite expansion of the distribution $f({\bf x},{\bf v},t)$ and
BGK collisional kernel \cite{shan_kinetic_2006}.

Knowledge of the discrete populations $f_{p}$ allows to compute the
local plasma density $\rho$, speed ${\bf u}$ and momentum-flux tensor
${\bf P}$, by a direct summation over the populations 
\begin{eqnarray}
\rho & = & \sum_{p}f_{p}\label{dens}\\
\rho{\bf u} & = & \sum_{p}f_{p}{\bf c}_{p}+\frac{h}{2}\rho{\bf G}\label{vel}\\
{\bf P} & = & \sum_{p}f_{p}{\bf c}_{p}{\bf c}_{p}\label{pre}
\end{eqnarray}
where eq. (\ref{vel}) ensures second order space/time accuracy of
the LB algorithm \cite{guo_discrete_2002}. 

The diagonal component of the momentum-flux tensor gives the plasma
pressure, while the off-diagonal terms give the shear stress $\boldsymbol{\sigma}$,
the latter being related to the non-equilibrium component of the populations,
\begin{equation}
\boldsymbol{\sigma}\equiv\nu\rho\left(\boldsymbol{\partial}{\bf u}+\boldsymbol{\partial}{\bf u}^{T}\right)=\frac{\nu}{c^{2}\tau}\sum_{p}{\bf c}_{p}{\bf c}_{p}\left(f_{p}-f_{p}^{eq}\right)\label{eq:stresstensorLB}
\end{equation}
On the other hand, due to the lack of a local kinetic definition of
the antisymmetric component of the displacement tensor $\partial{\bf u}$
and the fluid vorticity, these terms are evaluated via finite-differences.

\subsection{Red blood cells}

At first, red blood cells are introduced by modeling the hydrodynamic
interaction of a red blood cell with the surrounding plasma solvent
and next, by considering the collisional interaction of different
red blood cells among themselves. We take an oblate ellipsoid as a
good approximation to the hydrodynamic shape of a RBC.

A RBC is a body having mass $M$, position ${\bf R}_{i}$, velocity
${\bf V}_{i}$, angular velocity ${\bf \Omega}_{i}$, and instantaneous
orientation given by the matrix 
\begin{equation}
{\bf Q}_{i}=\left(\begin{array}{ccc}
\hat{n}_{x,i} & \hat{t}_{x,i} & \hat{g}_{x,i}\\
\hat{n}_{y,i} & \hat{t}_{y,i} & \hat{g}_{y,i}\\
\hat{n}_{z,i} & \hat{t}_{z,i} & \hat{g}_{z,i}
\end{array}\right)
\end{equation}
where $\hat{{\bf n}}_{i}$, $\hat{{\bf t}}_{i}$, $\hat{{\bf g}}_{i}$
are orthogonal unit vectors, such that ${\bf Q}_{i}^{T}{\bf Q}_{i}={\bf 1}$.
The orthogonal matrix ${\bf Q}_{i}$ transforms between the body and
the laboratory frame via ${\bf v}'={\bf Q}_{i}{\bf v}$, where the
primed and unprimed symbols stand for laboratory and body frames,
respectively. The tensor of inertia, ${\bf I}_{i}$, is diagonal in
the body frame and transforms to the laboratory frame according to
$\text{{\bf I}}'_{i}={\bf Q}_{i}\text{{\bf I}}_{i}\text{{\bf Q}}_{i}^{T}$.
In the sequel, we shall drop the prime symbol to ease the notation
and implicitly mean that the translational motion is handled in the
laboratory frame, where the rotational motion is handled in the body
frame.

We collectively denote the roto-translational state by the symbol
${\bf \Gamma}_{i}\equiv({\bf R}_{i},{\bf Q}_{i},{\bf V}_{i},{\bf \Omega}_{i})$.
Let us now introduce an auxiliary function to account for the shape
and orientation of the suspended RBC. We choose the following expression
\cite{peskin_immersed_2002} 
\begin{equation}
\tilde{\delta}({\bf x},{\bf Q}_{i})\equiv\prod_{\alpha=x,y,z}\tilde{\delta}_{\alpha}[({\bf {\bf Q}}_{i}{\bf x})_{\alpha}]
\end{equation}
with
\begin{equation}
\tilde{\delta}_{\alpha}(y_{\alpha})\equiv\left\{ \begin{array}{cc}
\frac{1}{8}\left(5-4|y_{\alpha}/\xi_{\alpha}|-\sqrt{1+8|y_{\alpha}|/\xi_{\alpha}-16y_{\alpha}^{2}/\xi_{\alpha}^{2}}\right)\;\;\;\; & |y_{\alpha}/\xi_{\alpha}|\leq0.5\\
\frac{1}{8}\left(3-4|y_{\alpha}|/\xi_{\alpha}-\sqrt{-7+24|y_{\alpha}|/\xi_{\alpha}-16y_{\alpha}^{2}/\xi_{\alpha}^{2}}\right)\;\;\;\; & 0.5<|y_{\alpha}/\xi_{\alpha}|\leq1\\
0 & |y_{\alpha}|/\xi_{\alpha}>1
\end{array}\right.
\end{equation}
and $\xi_{\alpha}$ being a set of three integers, one for each cartesian
component $\alpha=x,y,z$, representing the ellipsoidal radii in the
three principal directions. The shape function has compact support
and for $\xi_{x}=\xi_{y}=\xi_{z}=2$ generates a spherically symmetric
diffused particle with a support extending over $64$ mesh points.
Given that we handle one RBC via a single shape function, the computational
cost of a suspended RBC is proportional to the size of the support
in the three cartesian directions. 

The shape function has two important properties, it is normalized
when summed over the cartesian mesh points ${\bf x}$ \cite{peskin_immersed_2002}
\begin{equation}
\sum_{{\bf x}}\tilde{\delta}({\bf x}-{\bf T})=1
\end{equation}
for any continuous displacement ${\bf T}$, and obeys the property
\begin{equation}
\sum_{{\bf x}}(x_{\alpha}-T_{\alpha})\partial_{\beta}\tilde{\delta}({\bf x}-{\bf T})=-\delta_{\alpha\beta}\label{eq:intbypart}
\end{equation}

The translational response of the suspended body is designed according
to the RBC-fluid exchange kernel
\begin{equation}
\boldsymbol{\phi}({\bf {\bf x}},{\bf \Gamma}_{i})=-\gamma_{T}\tilde{\delta}({\bf x}-{\bf R}_{i},{\bf Q}_{i})\left[{\bf V}_{i}-{\bf u}({\bf x})\right]=-\gamma_{T}\tilde{\delta}_{i}\left({\bf V}_{i}-{\bf u}\right)\label{eq:pheno_force}
\end{equation}
where $\gamma_{T}$ is a translational coupling coefficient and where
the short-hand notation $\tilde{\delta}_{i}\equiv\tilde{\delta}({\bf x}-{\bf R}_{i},{\bf Q}_{i})$
has been introduced.

The body rotational response has different origins and can be analyzed
by considering the general decomposition of the deformation tensor
in terms of purely elongational and rotational terms 
\[
\boldsymbol{\partial}{\bf u}=\boldsymbol{e}+\boldsymbol{\rho}
\]
where ${\bf e}=\frac{1}{2}(\boldsymbol{\partial}{\bf u}+\boldsymbol{\partial}{\bf u}^{T})$
is the symmetric rate of strain tensor, related to the dissipative
character of the flow, and $\boldsymbol{\rho}=\frac{1}{2}(\boldsymbol{\partial}{\bf u}-\boldsymbol{\partial}{\bf u}^{T})$
is the antisymmetric vorticity tensor, which bears the conservative
component of the flow and is related to the vorticity vector $\boldsymbol{\omega}=\boldsymbol{\partial}\times{\bf u}=\boldsymbol{\epsilon}:\boldsymbol{\rho}$,
where $\boldsymbol{\epsilon}$ is the Levi-Civita tensor \cite{leal_advanced_2007}.
The rotational component of the deformation tensor gives rise to a
solid-like tumbling motion, where the rotational and elongational
one give rise to the vesicular, tank treading motion, as illustrated
in Fig. \ref{fig:flip-tank}. 

\begin{center}
\begin{figure}
\begin{centering}
\includegraphics[scale=0.8]{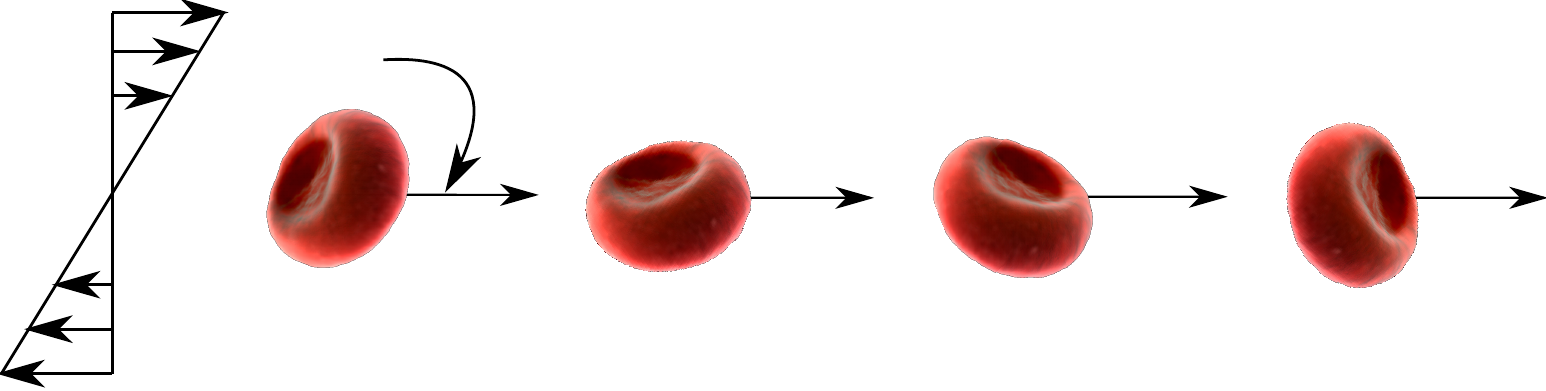}
\par\end{centering}

\centering{}\includegraphics[scale=0.8]{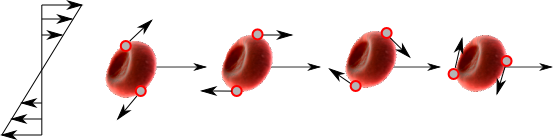}\caption{\label{fig:flip-tank} The two types of motion considered in modeling
a RBC in a linear shear field, the solid-like, tumbling or flipping
coin motion (upper panel), and the vesicular, tank-treading motion
(lower panel), where two material points on the RBC membrane move
at fixed body orientation.}
\end{figure}

\par\end{center}

Consequently, at rotational level, the RBC experiences two distinct
components of the torque. The first one arises from the coupling between
the body motion and the fluid vorticity, that we represent by the
following rotational kernel 
\begin{equation}
\boldsymbol{\tau}^{A}({\bf {\bf x}},{\bf \Gamma}_{i})=-\gamma_{R}\tilde{\delta}({\bf x}-{\bf R}_{i},{\bf Q}_{i})\left[{\bf \Omega}_{i}-{\bf \boldsymbol{\omega}}({\bf x})\right]=-\gamma_{R}\tilde{\delta}_{i}\left({\bf \Omega}_{i}-{\bf \boldsymbol{\omega}}\right)\label{eq:pheno_torque}
\end{equation}
where $\gamma_{R}$ is a rotational coupling coefficient and the superscript
$A$ stands for antisymmetric. This term depends on the body shape
and orientation via the shape function and, in a linear shear flow,
generates angular motion at constant angular velocity.

The elongational component of the flow contributes to the orientational
torque for bodies with ellipsoidal symmetry, being zero for spherical
solutes \cite{rioual_analytical_2004}. By defining the stress vector
${\bf t}^{\boldsymbol{\sigma}}=\boldsymbol{\sigma}\cdot\hat{{\bf n}}$,
where $\hat{{\bf n}}$ is the outward normal to the surface of a macroscopic
RBC, we replace the surface normal with the vector spanning over the
entire volume of the diffused particle, $\hat{{\bf n}}=\boldsymbol{\partial}\tilde{\delta}/|\boldsymbol{\partial}\tilde{\delta}|$.
The associated torque is represented in analogy with the torque acting
on macroscopic bodies \cite{leal_advanced_2007}, by the kernel 
\begin{equation}
\boldsymbol{\tau}^{S}({\bf {\bf x}},{\bf \Gamma}_{i})=\alpha\tilde{\delta}_{i}{\bf t}^{\boldsymbol{\sigma}}\times({\bf x}-{\bf R}_{i})\label{eq:pheno_torque_diss}
\end{equation}
where $\alpha$ is a parameter to be fixed and the superscript $S$
is mnemonic for the symmetric contribution of the flow. As shown in
the following, the elongational component of the torque includes an
independent contribution arising from tank treading that will allow
us to tune the parameter $\alpha$ based on known data on the tumbling
to tank treading transition.

The hydrodynamic force and torque acting on the RBC are obtained via
integration over the globule spatial extension. Owing to the discrete
nature of the mesh, the integrals are written as discrete sums, 
\begin{eqnarray}
{\bf F}_{i} & = & \sum_{{\bf x}}\boldsymbol{\phi}({\bf {\bf x}},{\bf \Gamma}_{i})=-\gamma_{T}({\bf V}_{i}-\tilde{{\bf u}}_{i})\\
{\bf T}_{i}^{A} & = & \sum_{{\bf x}}\boldsymbol{\tau}^{A}({\bf {\bf x}},{\bf \Gamma}_{i})=-\gamma_{R}({\bf \Omega}_{i}-\tilde{\boldsymbol{\omega}}_{i})\\
{\bf T}_{i}^{S} & = & \sum_{{\bf x}}\boldsymbol{\tau}^{S}({\bf {\bf x}},{\bf \Gamma}_{i})
\end{eqnarray}
where 
\begin{eqnarray}
\tilde{{\bf u}}_{i} & \equiv & \tilde{\delta}_{i}\star{\bf u}=\sum_{{\bf x}}\tilde{\delta}_{i}{\bf u}\label{eq:vel}\\
\tilde{\boldsymbol{\omega}}_{i} & \equiv & \tilde{\delta}_{i}\star\boldsymbol{\omega}=\sum_{{\bf x}}\tilde{\delta}_{i}\boldsymbol{\omega}\label{eq:vort}
\end{eqnarray}
are smeared hydrodynamic fields and the symbol $\star$ denotes convolution
over the mesh.

The action of the forces ${\bf F}_{i}$ and torques ${\bf T}_{i}={\bf T}_{i}^{S}+{\bf T}_{i}^{A}$
are counterbalanced by opposite reactions on the fluid side. Conservation
of linear and angular momentum in the composite fluid-RBC system preserves
the basic symmetries of the microdynamics and produces the consistent
hydrodynamic response \cite{duenweg_lattice_2008}. The action of
forces and torques on the fluid populations are expressed according
to the following body force 
\[
{\bf G}=-\sum_{i}\left\{ {\bf F}_{i}\tilde{\delta}_{i}+\frac{1}{2}{\bf T}_{i}\times\boldsymbol{\partial}\tilde{\delta}_{i}\right\} 
\]
The two exchange terms arising from the translational and rotational
back-reactions produce distinct modifications of the fluid velocity
and vorticity. Some algebra shows that every suspended body preserves
mass and linear momentum in the composite fluid-RBC system since $\sum_{{\bf x}}\Delta{\bf u}=\sum_{{\bf x}}\sum_{p}\Delta f_{p}{\bf c}_{p}=-{\bf F}_{i}$.
Similarly, every suspended body preserves the total angular momentum,
since 
\begin{equation}
\sum_{{\bf x}}\Delta\boldsymbol{\omega}=\sum_{{\bf x}}\sum_{{\bf p}}\Delta f_{p}{\bf c}_{p}\times({\bf x}-{\bf R}_{i})=\frac{1}{2}\sum_{{\bf x}}\left({\bf T}_{i}\times\boldsymbol{\partial}\tilde{\delta}_{i}\right)\times({\bf x}-{\bf R}_{i})=-{\bf T}_{i}
\end{equation}
where we have used the Lagrange rule, ${\bf a}\times({\bf b}\times{\bf c})=({\bf a}\cdot{\bf c}){\bf b}-({\bf a}\cdot{\bf b}){\bf c}$
and the property of the shape function, eq. (\ref{eq:intbypart}).

\subsection{Tank Treading}

Tank Treading (TT) is the motion featured by vesicles and RBC and
arises from the sliding of the surrounding membrane when subjected
to a shearing flow. It can be visualized by considering a material
point anchored to the RBC membrane and moving in elliptical orbits
while maintaining fixed the orientation of the RBC with respect to
the shearing direction (see Fig. \ref{fig:flip-tank}). The physical
parameters controlling tank treading are the ratio of viscosity between
plasma and the inner fluid entrained within the RBC, the shape of
the RBC and the shear rate $\dot{\gamma}$ \cite{keller_motion_1982}. 

In modeling blood, it is essential to include TT as it acts to orient
RBC with a privileged angle with respect to the flow direction. In
proximity of the vessel walls, the fixed orientation induces a net
lift force proportional to $\dot{\gamma}$ that pushes the globule
away from the walls \cite{olla_simplified_1999}. Lift forces, arising
from either single-body or many-body effects, are thought to induce
the Farhaeus-Lindqvist phenomenon, the drop of blood viscosity in
vessels of sub-millimeter diameters, an effect with far-reaching consequences
in physiology \cite{lightfoot_transport_1974}. 

In the general case, TT motion takes place in an apparently decoupled
fashion from tumbling, as vorticity and elongational contributions
have different origins. However, the motion of the rigid-body RBC
is partially compensated by the movement of the surrounding membrane
and thus effectively couples tumbling and TT motion via the instantaneous
orientation of the globule. At small values of the viscosity ratio,
orientational torques prevail over the rotational ones and pure TT
motion with a fixed RBC orientation is observed. The phase diagram
as a function of the viscosity ratio provides a direct view on the
RBC response to shear \cite{kessler_swinging_2008}.

An analytical treatment accounting for tank treading is due to Skalak
and Keller (SK) \cite{keller_motion_1982} a model that has enjoyed
a large popularity in determining the single-body forces and torques
acting on RBCs \cite{messlinger_dynamical_2009} %
\footnote{In its original formulation, the Skalak-Keller model is two-dimensional
and considers the torques acting on elliptical RBC in Stokes flow
and obeying the superposition principle, that is, with tumbling and
tank treading taken as independent motions. Moreover, the RBC is considered
as passive scalar with no back-reaction on the fluid. The SK model
predicts tumbling to TT transition independent on the shear rate $\dot{\gamma}$. %
}. Recently the SK model has been re-derived by Rioual et al. \cite{rioual_analytical_2004}
via a heuristic argument that we recall here. In 2D, when the RBC
is subjected to a velocity field ${\bf u}=(0,\dot{\gamma}y)$, the
elongational component tends to align the RBC along the shearing principal
eigenvector. For an elliptical body, the alignment has an angle $\psi=\frac{\pi}{4}$,
where $\psi$ is the angle formed by the RBC largest principal axis
with respect to the flow direction.

In general, the torque acting on a RBC can be decomposed into three
separate components. The first component is due to the angular motion
of RBC in a quiescent fluid and produces a frictional torque given
by $-\gamma_{R}\boldsymbol{\Omega}$, where $\gamma_{R}$ is a phenomenological
coefficient that can be identified with the one introduced in eq.(\ref{eq:pheno_torque}).
The second component is due to the torque from the shearing fluid
on the quiescent RBC and for a quiescent membrane, being related to
the vorticity component of the flow. The analytical solution for the
2D elliptical RBC without TT motion provides a total torque proportional
to $\dot{\gamma}\left(\frac{L_{1}^{2}+L_{2}^{2}}{2}+\frac{L_{1}^{2}-L_{2}^{2}}{2}\cos2\psi\right)$,
where $L_{1,2}$ are two shape-dependent coefficients and where the
first term of the expression is associated to vorticity while the
second term to the elongational component of the flow. The latter
provides two equilibrium positions for $\psi=\pm\frac{\pi}{4}$ and
maximal torque for $\psi=0,\frac{\pi}{2}$. 

Finally, the effect of tank treading is determined by considering
the local frame moving together with the material point at velocity
${\bf V}_{TT}$ and with the infinitesimal membrane element experiencing
a force $d{\bf F}_{TT}\propto{\bf V}_{TT}$ and torque $\boldsymbol{\tau}_{TT}=\oint d{\bf F}_{TT}\times({\bf x}-{\bf R})$.
Consequently, TT couples to both the rotational and elongational flow
components and results in a net torque $\boldsymbol{\tau}_{TT}\propto\cos(2\psi)$,
that is, a torque enjoying the same angular symmetry of the mechanism
associated to the rigid body response. Given the three-dimensional,
diffused nature of the numerical model, the actual presence of the
cellular membrane is not explicitly considered. Instead, the effect
of TT is controlled by tuning the intensity of the elongational torque
via the adimensional factor $\alpha$ of eq. (\ref{eq:pheno_torque_diss}).

\subsection{Viscosity contrast}

RBC are carriers of an internal fluid composed of several hundred
millions of hemoglobin proteins and being much more viscous than plasma.
The inner fluid contributes significantly to the dissipation of energy
as mediated by the shearing modes activated by the RBC membrane. The
consequence is the steep raise in the apparent viscosity with the
hematocrit level. We incorporate the effect of the viscosity contrast
by considering a local enhancement of the LB fluid viscosity within
the RBC shape according to the following BGK relaxation time
\begin{equation}
{\cal T}(x)={\cal T}_{0}+\Delta h\sum_{i}\tilde{\theta}_{i}\label{eq:enhance-viscosity}
\end{equation}
where ${\cal T}_{0}$ corresponds to the viscosity of pure plasma,
$\Delta$ is viscosity enhancement factor, and $\tilde{\theta}_{i}$
is a smooth version of the ellipsoidal characteristic function. The
latter is represented as $\tilde{\theta}_{i}=1-(1-\tilde{\delta_{i}})^{\kappa}$
where the parameter $\kappa$ governs the smooth transition between
the inner and outer fluids, chosen as $\kappa=20$. By choosing $\nu_{0}=1/6$
and $\Delta=2$, the ratio between inner ($\tilde{\theta}_{i}\sim1$)
and outer ($\tilde{\theta}_{i}\sim0$) viscosities is equal to $5$.

\subsection{Excluded Volume Interactions}

In order to account for the RBC-RBC repulsive forces, soft-core mechanical
forces and torques are given by the Gay-Berne (GB) potential \cite{gay_modification_1981},
as widely used in the liquid crystals community. The GB potential
inhibits interpenetration of pairs of RBCs by introducing an orientation-dependent
repulsive interaction derived from the Lennard-Jones potential ($\phi_{LJ}(R_{ij})=\epsilon[(\sigma/R_{ij})^{12}-(\sigma/R_{ij})^{6}]$,
where $\epsilon$ is the energy scale and $\sigma$ the ``contact''
length scale). The GB potential extends the spherically symmetric
Lennard-Jones potential to ellipsoidal symmetry, where the potential
depends if two RBCs have mutual orientation as face-to-face (maximal
repulsion), side-to-side (minimal repulsion) or an arbitrary orientation
between the two bodies (intermediate case). 

In the following, we shall employ the form of the GB potential that
handles the interaction between particles of different eccentricity,
such as a mixture of ellipsoidal and spherical particles. This flexibility
allows to handle biofluids composed of particles with different shapes
by employing the same analytical form of the potential.

Given the principal axes $(a_{i,1},a_{i,2},a_{i,3})$ of the $i$-th
globule, the ellipsoidal shape associated to the excluded volume interactions
is constructed according to the shape matrix $S_{i}=\mbox{{diag}}(a_{i,1},a_{i,2},a_{i,3})$
and the transformed matrix ${\bf A}_{i}={\bf Q}_{i}{\bf S}_{i}^{2}{\bf Q}_{i}^{T}$
in the laboratory frame. The pair of particles $i,j$ at distance
${\bf R}_{ij}$ experiences a characteristic exclusion distance $\sigma_{ij}$
that depends on the globule-globule distance, shape and mutual orientation,
written as
\begin{eqnarray}
\sigma_{ij} & = & \frac{1}{\sqrt{\phi_{ij}}}\label{eq:sigmagb}\\
\phi_{ij} & = & \frac{1}{2}\hat{{\bf R}}_{ij}\cdot{\bf H}_{ij}^{-1}\cdot\hat{{\bf R}}_{ij}\label{eq:phigb}
\end{eqnarray}
where the matrix ${\bf H}_{ij}={\bf A}_{i}+{\bf A}_{j}$ has been
introduced. 

A purely repulsive exclusion potential is given by the pairwise form
\cite{gay_modification_1981,allen_expressions_2006} 
\begin{equation}
u_{ij}=\left\{ \begin{array}{lc}
4\epsilon_{0}(\rho_{ij}^{-12}-\rho_{ij}^{-6})+\epsilon_{0} & \qquad\rho_{ij}^{6}\leq2\\
0 & \qquad\rho_{ij}^{6}>2
\end{array}\right.\label{eq:gayberne}
\end{equation}
with
\begin{equation}
\rho_{ij}=\frac{R_{ij}-\sigma_{ij}+\sigma_{ij}^{min}}{\sigma_{ij}^{min}}
\end{equation}
where $\epsilon_{0}$ is the energy scale and $\sigma_{ij}^{min}$
is a constant, both parameters being independent on the ellipsoidal
mutual orientation and distance. For two identical oblate ellipsoids,
$\sigma_{ij}^{min}$ corresponds to a contact distance of the two
particles having face-to-face orientation. In general, by considering
the minimum particle dimension $a_{i}^{min}=\min(a_{i,1},a_{i,2},a_{i,3})$
then 
\begin{equation}
\sigma_{ij}^{min}=\sqrt{2\left[\left(a_{i}^{min}\right)^{2}+\left(a_{j}^{min}\right)^{2}\right]}
\end{equation}
The mutual forces and torques exerted between the $i,j$ pair are
computed accordingly, as described in the appendix for the sake of
completeness.

We note here two aspects. First, the softness of the RBC can be improved
by taking small values for the energy scale $\epsilon_{0}$. Clearly,
this type of softness does not arise from hydrodynamic forces but
from RBC-RBC mechanical forces, and emulates the good packing attitude
of RBCs at high hematocrit levels. Secondly, the ellipsoidal exclusion
shape can be modified to reproduce more closely the biconcave discoid
via the Bates-Lockhurst version of the GB potential \cite{care_computer_2005}.
At present, we take the ellipsoidal shape to be representative of
RBC, while further refinements can be trivially included. 

When simulating blood flows in a confining environment, we associate
to each wall mesh point a spherical particle that acts as a repulsive
center for the ellipsoidal RBC and prevents the globule from leaking
out of the confining vessel, under the action of the same GB potential.
The wall-RBC characteristic distance $\sigma_{w,RBC}^{min}$ is chosen
to be half the mesh spacing $\frac{\Delta x}{2}$.

\subsection{Summary of the computational model}

To recap, the basic assumption of the model is to handle a suspension
made of plasma and RBC. The former is a Newtonian fluid that reproduces
Navier-Stokes dynamics at virtually arbitrary Reynolds numbers and
in arbitrary geometries. The physiological conditions of Reynolds
$\lesssim2000$ and shear rates $\lesssim500\, s^{-1}$ can be accessed
in simulation without posing limitations in terms of numerical robustness
and feasibility. The model for RBCs has the following characteristics:
\begin{itemize}
\item it is based on the body forces and torques acting on the RBC rather
than on surface forces by a proper decomposition of the translational,
tumbling and orientational components of the flow. A similar scenario
is found in Faxen's law for zero-Reynolds flows \cite{leal_advanced_2007}.
Our strategy does not track explicitly the RBC membrane or handle
in anyway the internal cytoskeleton;
\item RBCs are active scalars with hydrodynamic shape that is fixed and
ellipsoidal. The eccentricity of the RBC can be tuned, however in
the following we choose to work with $\xi_{x}=1$ and $\xi_{y}=\xi_{z}=2$,
corresponding to volume $V\simeq134$, surface $S\simeq139$ and reduced
volume $v\equiv\frac{3V}{4\pi(S/4\pi)^{3/2}}\simeq0.87$, that should
be compared with $v=0.65$ for human RBCs;
\item The near-field hydrodynamic interactions are well captured by the
model, in particular as regarding to translational and orientational
mobilities and long-time tails arising from hydrodynamic interactions.
Shape fluctuations can be easily introduced, e.g. by means of the
Maffettone-Minale extension \cite{maffettone_equation_1998}, by allowing
volume-preserving deformations while still maintaining the ellipsoidal
symmetry;
\item the viscosity contrast is introduced by allowing for a modification
of the relaxation time within the RBC region. In practice, this means
solving the same LB algorithm throughout the system, but with a local
relaxation time and thus, higher dissipation within the RBC region.
In terms of computational feasibility, a wide range of viscosity contrast
can be explored, but in the following we will focus on a given viscosity
contrast as pertaining to RBCs;
\item interpenetration between RBCs is avoided by Gay-Berne forces and torques.
The degree of softness of particles can be tuned, as much as some
attracting component of the interaction due to adhesive forces or
the presence of fibrinogen, can be easily accommodated within the
model.
\end{itemize}
In summary, the main assumption of the model, the neglect of shape
fluctuations, is largely compensated by its main strength, the possibility
to handle large-scale systems of physiological relevance with state-of-the-art
computer hardware. This is a major point of the model that will not
be addressed in detail here, but has been discussed in refs.\cite{bernaschi_muphy:_2009,bernaschi_flexible_2010,peters_multiscale_2010}.

\section{Numerical Results \label{sec:Numerical-Results}}

The frictional response of a single RBC in plasma is first analyzed
by computing the Stokes response of an oblate ellipsoidal particle
having two possible orientations with respect to the motion direction.
For translational displacement, these are the frontal and side-wise
motions. For the rotational motion, these correspond to spinning around
two principal directions corresponding to the smallest and largest
radii. For this preliminary test, we deactivate the modulation of
viscosity by the presence of the RBC, that is, we take the same viscosity
for plasma and the entrained fluid without distinction and throughout
the simulation domain.

As shown in ref. \cite{ahlrichs_simulation_1999} for a model of suspended
point-like particles, the effective mobility is given by the sum of
two components, the mobility associated to the bare frictional parameters,
$\gamma_{T}$ and $\gamma_{R}$, and the effect of the hydrodynamic
field induced on the surrounding solvent that sustains the motion
by increasing the particle roto-translational mobilities. In addition,
the hydrodynamic components to mobility contains a Stokes-like component
that is renormalized by the presence of the numerical finite-spacing
mesh. We expect to recover the same qualitative behavior for the diffused
ellipsoidal particles.

Fig.\ref{fig:mobility} shows the computed particle mobilities as
a function of the frictional parameters $\gamma_{R}$ and $\gamma_{T}$
for a RBC of mass $M=10$ and inertia $I_{x,y,z}=1000$ (all data
are expressed in lattice units unless otherwise stated). The particle
is embedded in a simulation box of size $60\times60\times60$ containing
a quiescent fluid. At infinite friction, the intercepts correspond
to the mesh-induced spurious frictional forces. For translational
motion, the intercepts correspond to two mobilities, $\mu\equiv\frac{1}{M\gamma_{T}^{mesh}}=6.6$
and $10.3$, for frontal and side-wise translation motion, respectively.
By associating a residual hydrodynamic radius $a_{T}^{mesh}=\frac{M\gamma_{T}^{mesh}}{6\pi\nu\rho}=0.06$
and $0.03$ to frontal and side-wise motion, respectively, it is apparent
that the residual radius is directly proportional to the size factor
$\xi_{\alpha}$ governing the shape function. From these data it is
clear that the residual Stokes radii are much smaller than the mesh
spacing (being $\Delta x=1$ in lattice units) and modulating the
hydrodynamic response to achieve a bulkier suspended body requires
increasing the coupling parameter to $\gamma_{T}>10$. High values
of $\gamma_{T}$ spoil the numerical robustness of the method. In
fact, a simple estimate shows that numerical stability of the LB method
is for body forces $|F|\lesssim0.1$, that is $\gamma_{T}\lesssim0.1$.
The translational data show that the coupling method alone reproduces
the far-field Stokes response but is unable to modulate the viscosity
to significant values with the hematocrit level, unless a different
mechanism is introduced to improve the internal dissipation, such
as the one encoded by eq. (\ref{eq:enhance-viscosity}).

For angular motion, the situation is slightly different since we find
the intercepts $\gamma_{R}^{mesh}=\frac{1}{44}\,\Delta t^{-1}$ and
$\frac{1}{38}\,\Delta t^{-1}$ for rotations around the smallest and
largest principal radii, respectively. The intercepts correspond to
the residual rotational radii $a_{R}^{mesh}=\left(\frac{I\gamma_{R}^{mesh}}{8\pi\nu\rho}\right)^{1/3}=1.84\,\Delta x$
and $1.75\,\Delta x$, respectively. In the angular motion, the mesh-induced
friction is larger than the translational counterpart (in fact, comparable
to the mesh spacing) and exhibits a weak dependence on the direction
of spinning direction, so that it can be considered independent on
the latter. 

\begin{center}
\begin{figure}
\centering{}\includegraphics[scale=0.5]{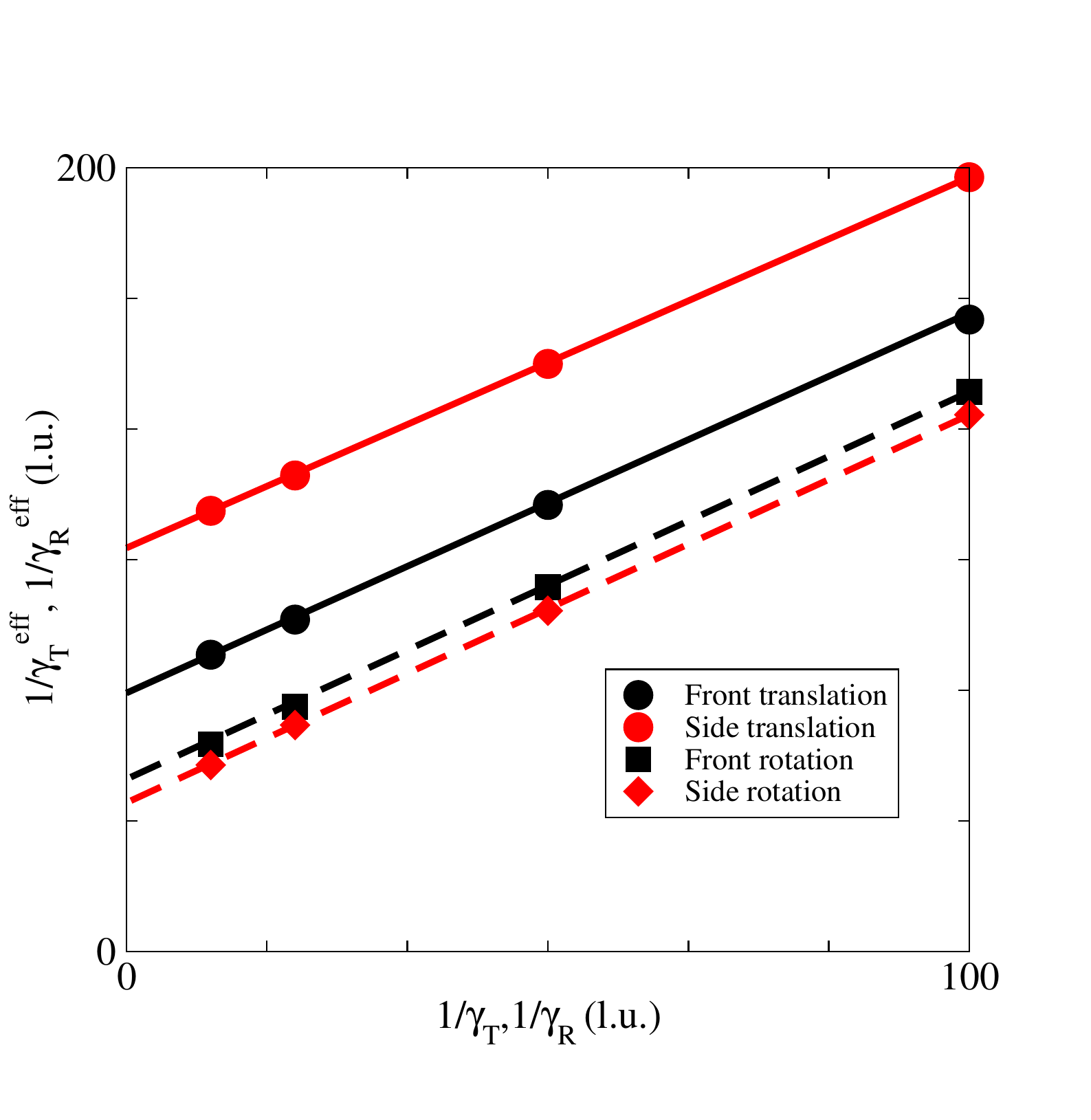}\caption{\label{fig:mobility}Translational and rotational mobilities as a
function of the coupling parameters $\gamma_{T}$ and $\gamma_{R}$.
Circles correspond to frontal (filled symbols) and lateral translation
(open symbols). Squares are for rotations around the smallest principal
radius (filled symbols) and around the largest principal radius (open
symbols). The lines are guides for the eye. }
\end{figure}

\par\end{center}

The effect of tank treading is analyzed by placing a RBC in a shear
field generated according to the method of ref. \cite{wagner_lees-edwards_2002}.
Fig. \ref{fig:TT-torque} shows the time dependence of the elongational
component of the torque for a single RBC in a shear flow. Different
values of the enhancement factor $\alpha$ are reported having chosen
$\gamma_{T}=\gamma_{R}=10^{-1}$. It is found that the value $\alpha=0.063$
corresponds to the transition to pure TT motion, where the elongational
and vorticity-related torques exactly compensate, generating a constant
orientation with angle $\simeq40^{o}$ as a fixed point of the dynamics.
For a bounded fluid in a two-dimensional slab the fixed angle is found
to be in the range $20^{o}-35^{o}$, depending on the viscosity contrast
and flow curvature, with a lift force being linearly proportional
to the local shear rate and to $y^{-2}$, where $y$ is the distance
of the RBC from the wall, in agreement with experimental and theoretical
results \cite{cantat_lift_1999,coupier_noninertial_2008}. 

Overall, the torques generate a well-behaved rotational motion and
in line with the expected trend. In the following we arbitrarily set
the tunable parameter $\alpha=0.04$, that is, slightly below the
transition to tank treading, and proceed further with tests regarding
the rheological response of a dense suspension. The tests in the bounded
fluid confirm the tumbling regime for the chosen RBC shape and viscosity
ratio. It should be borne in mind that the balance between tumbling
and tank treading in a dilute suspension critically affects the lift
forces and thus the viscosity of the suspension \cite{danker_rheology_2007}.
In this respect, the fixed-shape approximation of the present model
produces hydrodynamic forces that are qualitatively correct. In the
future, the model can be tuned more finely by optimizing the shape
and viscosity contrast.

\begin{center}
\begin{figure}
\centering{}\includegraphics[scale=0.5]{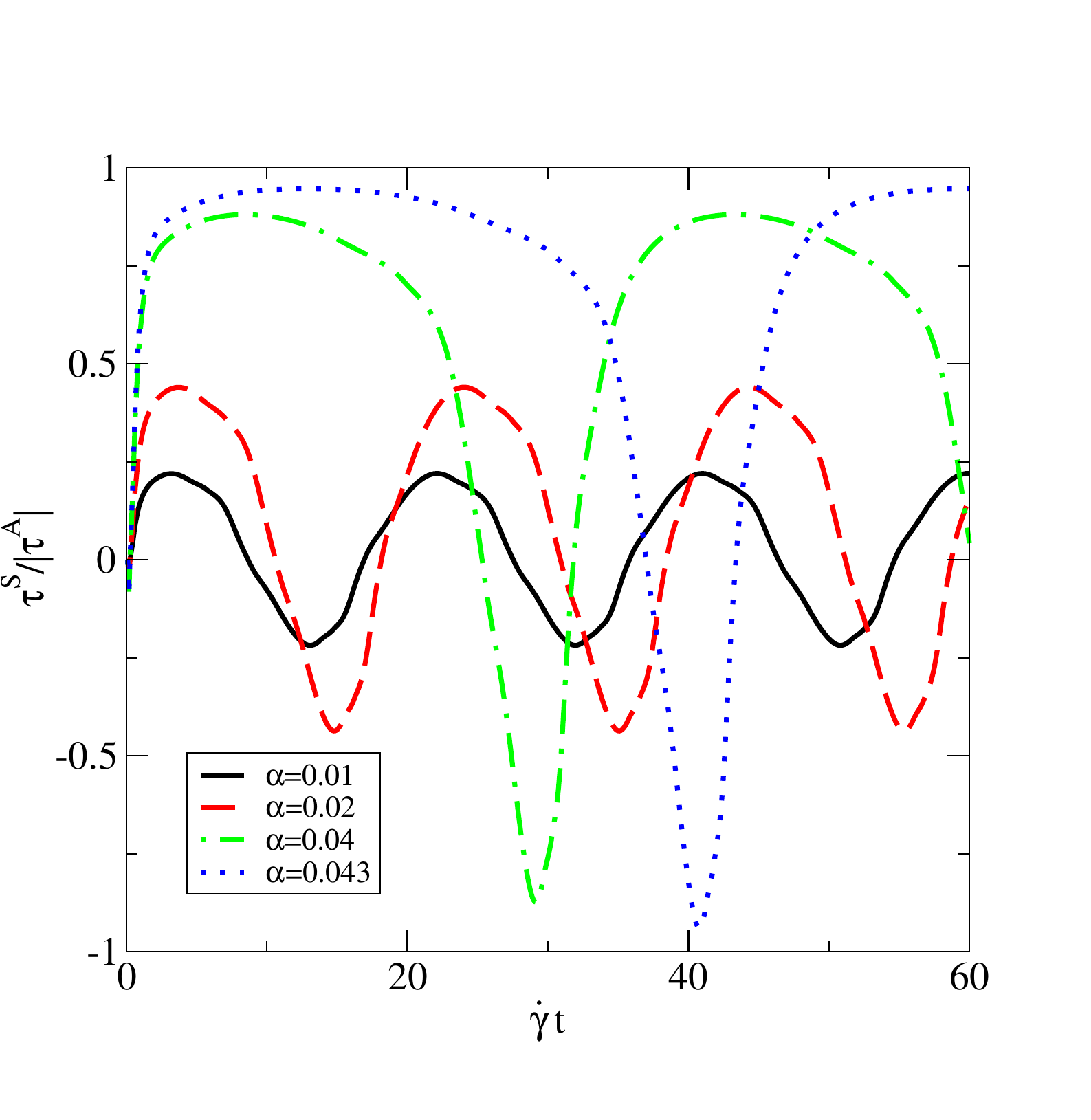}\caption{\label{fig:TT-torque}Symmetric component of the torque of a single
RBC in a shear field and for different values of the tank treading
parameter $\alpha$. }
\end{figure}

\par\end{center}

The viscosity of the suspension is computed by using a cylindrical
channel of radius $50\,\mu m$ and for different hematocrit levels.
In Fig. \ref{fig:appvisc-channel} we report the relative viscosity
($\eta_{rel}=\eta_{app}/\eta_{0})$ where $\eta_{app}$ is the apparent
viscosity measured in the channel at finite hematocrit and $\eta_{0}$
the viscosity in the same channel at zero hematocrit. Since $\eta\propto Q$,
where $Q$ is the volumetric flow rate, $\eta_{rel}$ is computed
as the ratio of flow rates at finite and zero hematocrit. Fig. \ref{fig:appvisc-channel}
also reports the data on viscosity by setting the enhancement factor
of eq. (\ref{eq:enhance-viscosity}) to $\Delta=0$. The latter produce
a weak modulation of viscosity with hematocrit, while the data with
$\Delta=2$ exhibit an excellent agreement with the experimental results
of ref. \cite{pries_blood_1992}.

\begin{center}
\begin{figure}
\centering{}\includegraphics[scale=0.5]{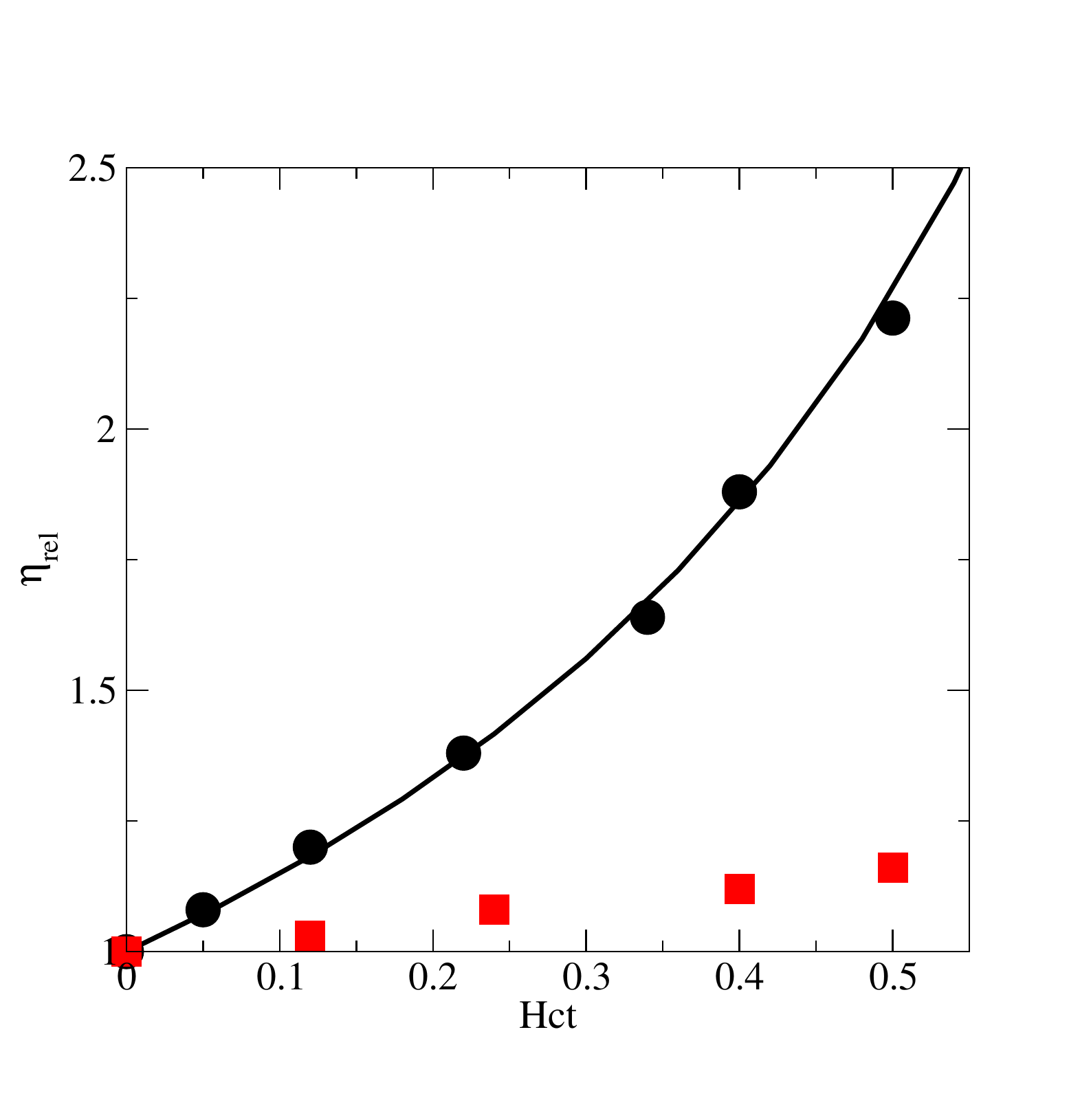}\caption{\label{fig:appvisc-channel}Relative viscosity in a channel of radius
$50\,\mu m$ for different hematocrit levels as compared to the experimental
data of Pries et al. \cite{pries_blood_1992}(solid curve). Data are
for the enhanced dissipation mechanism of eq. (\ref{eq:enhance-viscosity})
with $\Delta=2$ (circles) and without enhancement ($\Delta=0$) (squares).}
\end{figure}

\par\end{center}

A crucial phenomenology of blood circulation is the decrease of viscosity
as the vessel radius falls below $100\,\mu m$, namely, the Farhaeus-Lindqvist
effect \cite{lightfoot_transport_1974}. This effect is usually ascribed
to the formation of a cell-free layer in proximity of the vessel walls.
The origin of such depletion is still uncertain but the lateral forces
that push the RBCs away from the vessel walls are retained to have
different causes, such as tank treading and cell deformation \cite{olla_simplified_1999},
adhesive properties of RBC or shear-induced migration. In the current
version of our model, we do not probe the effects of cell deformation.
However, the simulation reveals a distinct RBC depletion in proximity
of the walls, as shown in Fig. \ref{fig:cellfreelayer}. We report
the variation of the cell-free layer with the hematocrit and vessel
diameter as compared to the experimental data of Bugliarello and Sevilla
\cite{bugliarello_velocity_1970}. The numerical results reproduce
the experimental data quite well, lending good confidence in the numerical
model at vessel diameters below the $100\,\mu m$ radius. The typical
cell-free layer observed in simulation for two channels of different
radii is illustrated in Fig. \ref{fig:cellfreelayer}, inset. The
progressive depletion of RBC as compared to the plasma content is
the leading cause of fluidization of the suspension at small channel
radius.

\begin{center}
\begin{figure}
\centering{}\includegraphics[scale=0.5]{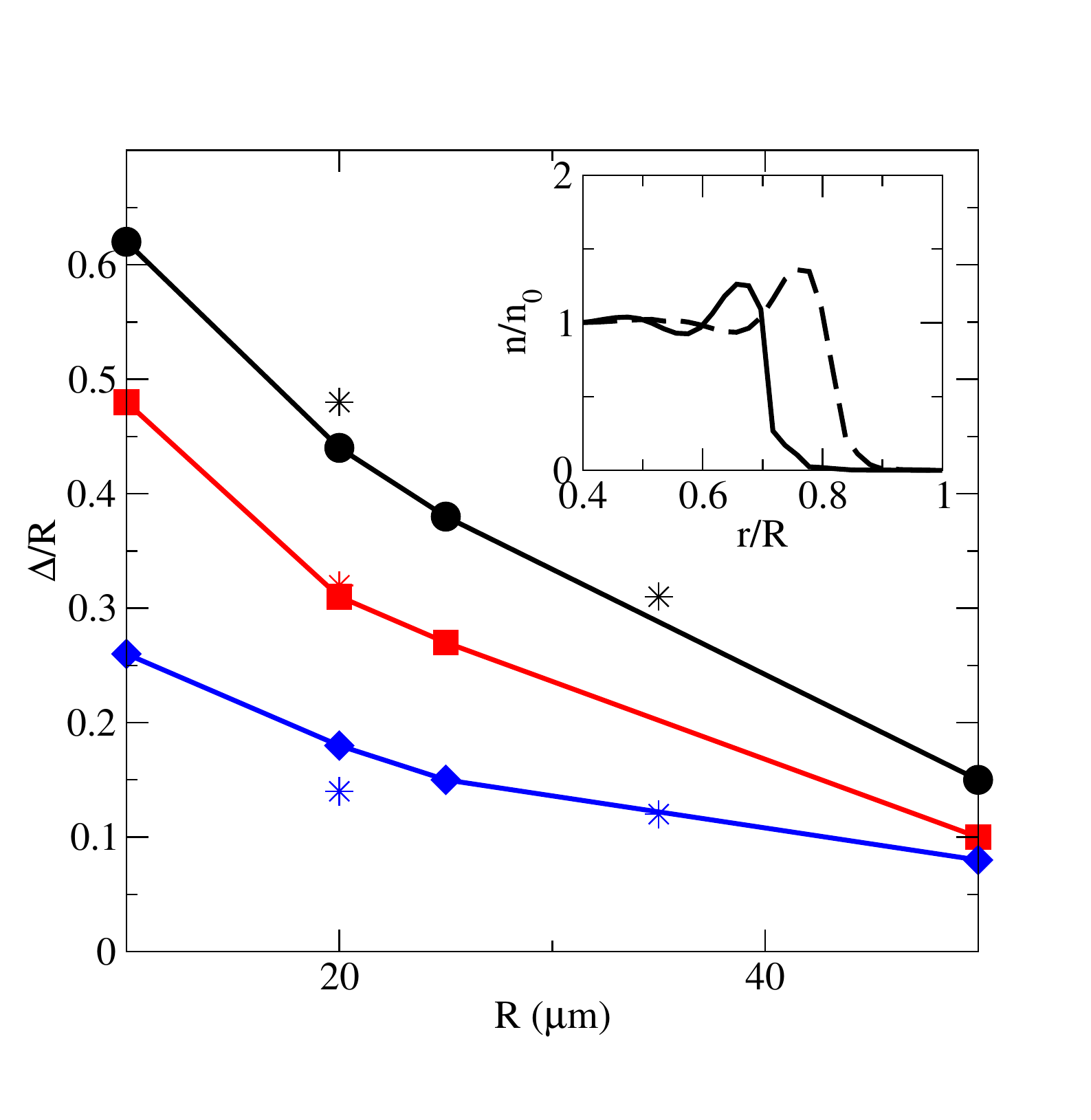}\caption{\label{fig:cellfreelayer}Size dependence of the cell-free layer with
the vessel diameter and hematocrit level of $10\,\%$ (diamonds),
$20\,\%$ (squares) and $50\,\%$ (circles), as compared to the experimental
data of Bugliarello and Sevilla \cite{bugliarello_velocity_1970}
(star symbols). The lines are guides for the eye. Inset: Radial density
profiles of RBC, for $R=10\,\mu m$ (solid line) and $R=20\,\mu m$
(dashed line), illustrating the cell-free layers in proximity of the
vessel wall.}
\end{figure}

\par\end{center}

The non-Newtonian behavior of the suspension is further exhibited
by the velocity profiles of RBC for different hematocrit levels and
vessel diameters, as shown in Fig. \ref{fig:velprofiles}. As the
hematocrit level increases, the Poiseuille-like parabola modifies
into flatter profiles next to the vessel centerline and in a large
extension of the channel, whereas in proximity of the walls, the profiles
have large slopes and strong dissipation, in particular for the narrower
vessels. We further notice a good match of the plasma and RBC velocities
for all vessel radii, a matching that is typically lost as the vessel
radius becomes comparable to the RBC size (data not shown).

\begin{center}
\begin{figure}
\centering{}\includegraphics[scale=0.6]{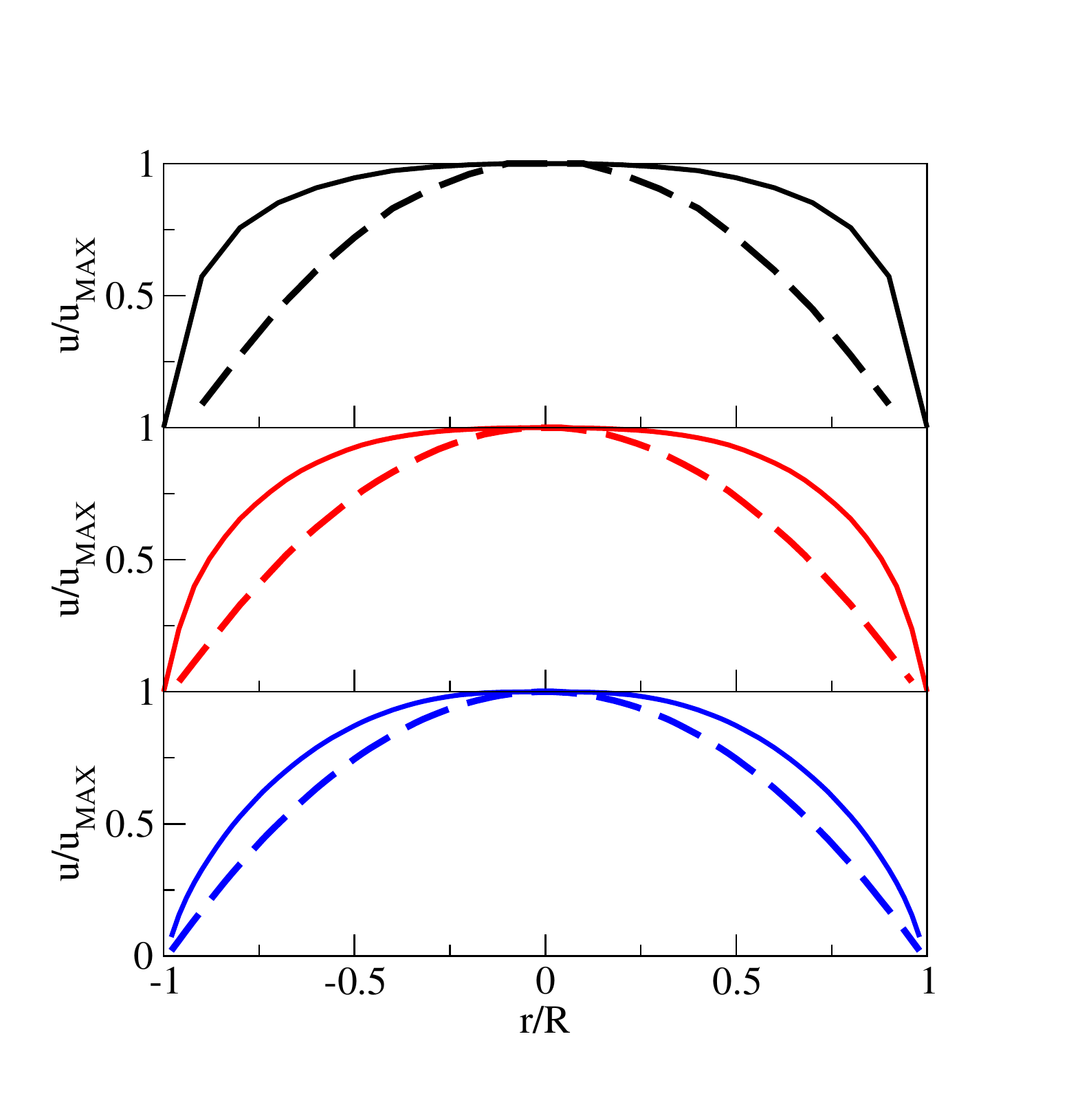}\caption{\label{fig:velprofiles}Velocity profiles for vessel radius of $10$
(upper panel), $25$ (mid panel) and $50\,\mu m$ (lower panel). Data
correspond to hematocrit levels of $35\%$ (solid lines) and $0\%$
(dashed lines). }
\end{figure}

\par\end{center}

In applying the model to physiological conditions, we consider a realistic
bifurcating vessel at $50\%$ hematocrit level, as depicted in Fig.
\ref{fig:bifurca}. The bifurcation is extracted as part of a coronary
arterial system \cite{melchionna_hydrokinetic_2010}, and is made
of a parent vessel of radius $\sim100\,\mu m$ and one daughter branch
having approximately the same size of the parent vessel, and a second
daughter branch of radius $\sim80\,\mu m$. 

The flow is induced by imposing parabolic flow conditions at the inlet
and constant pressure at the outlets, with the method described in
\cite{peters_multiscale_2010}, producing a shear rate of $80\, s^{-1}$
averaged over the entire bifurcation at steady state, and reaching
as high as $150\, s^{-1}$ in proximity of the arterial wall, values
in line with the typical shear rates observed in human coronary arteries,
typically in the $30-450\, s^{-1}$ range \cite{ethier_introductory_2007}.
As the snapshot reveals, RBCs organize in several different ways throughout
the bifurcation and also depending on the value of the shear rate
(data not shown). The local organization of RBCs in rouleaux is clearly
visible, the typical stack often observed in static or flow conditions,
and mostly destroyed as the shear rate increases.

\begin{center}
\begin{figure}
\centering{}\includegraphics[scale=0.5]{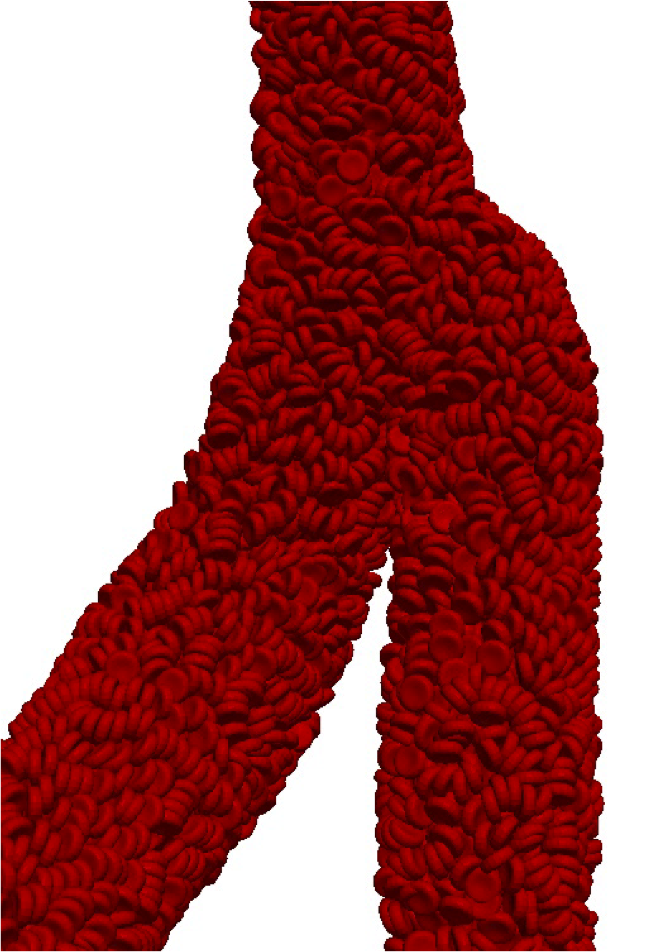}\caption{\label{fig:bifurca}Snapshot of the bifurcating vessel showing the
organization of RBCs at hematocrit of $50\,\%$ and average shear
rate of $80\, s^{-1}$.}
\end{figure}

\par\end{center}

\begin{figure}
\centering{}\includegraphics[scale=0.5]{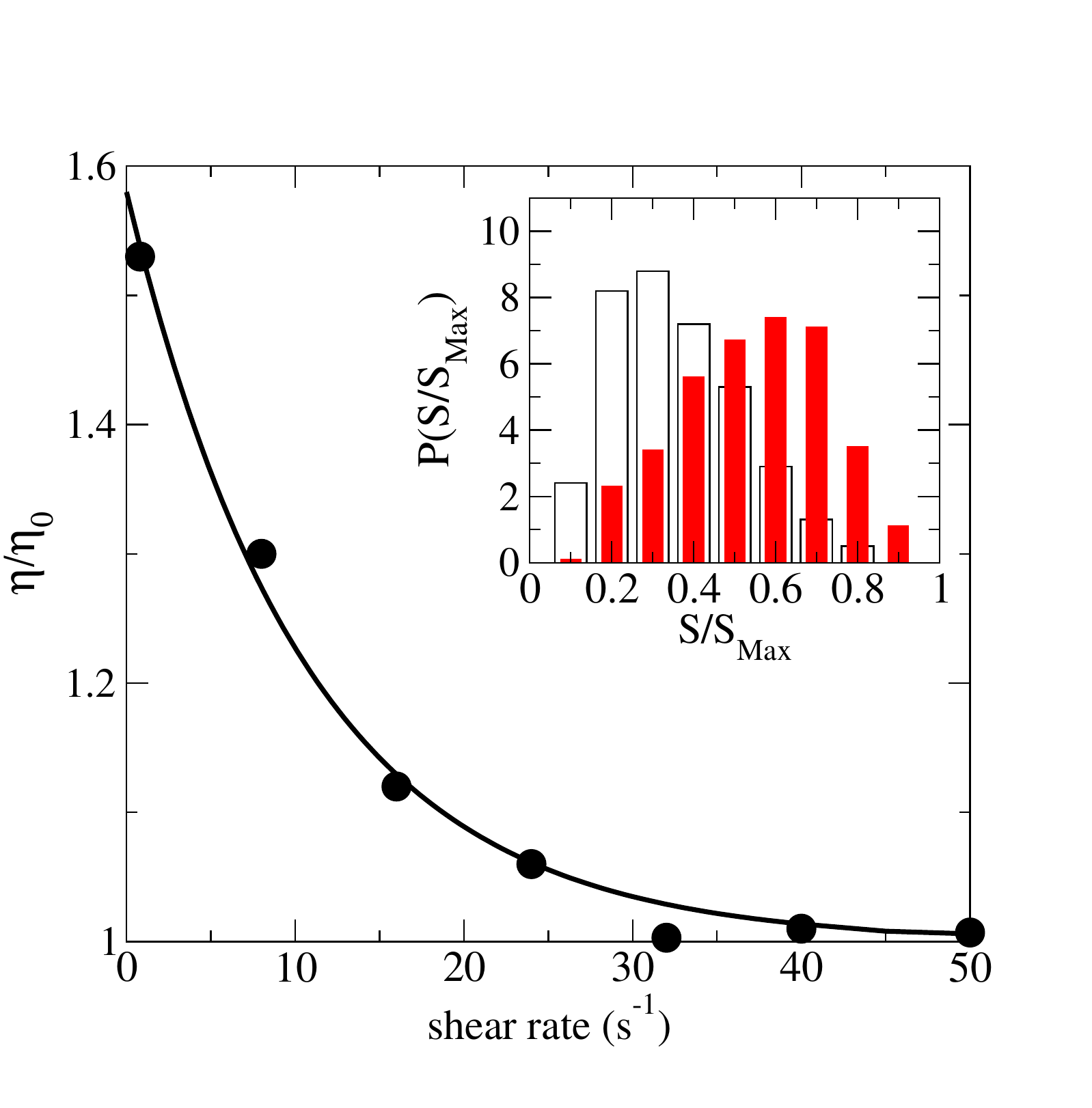}\caption{\label{fig:shearthinning}Shear thinning of the suspension in the
bifurcating channel for different values of the shear rate. Inset:
Distribution of rouleaux as a function the cluster size $S$, for
shear rate of $8\, s^{-1}$ (solid bars) and $16\, s^{-1}$ (open
bars). }
\end{figure}

The flow in the bifurcating channel for different values of the shear
rate is analyzed and the shear thinning behavior is illustrated in
Fig. \ref{fig:shearthinning}, where viscosity decreases with the
shear rate until the Newtonian plateau is reached. 

As expected, shear thinning correlates with the formation of rouleaux,
as seen by inspecting the statistical distribution of cluster size.
The histogram of the rouleaux size is reported in Fig. \ref{fig:shearthinning},
inset. As shear rate increases, fewer and smaller structures are detected.
However, a central observation is that the local structuring of RBCs
strongly depends on the morphological details of the vessel. In fact,
we notice that in proximity of the bifurcation a shoulder in a daughter
vessel creates smaller flow velocity and some stagnation of RBCs.
In correspondence of this region, rouleaux show resilient character
even for shear rates above $50\, s^{-1}$. 

The uneven distribution of RBCs, with the attendant stagnation and
persistence of rouleaux in specific regions, can have significant
impact on the distribution of shear stress. The latter quantity strongly
correlates with the formation of plaques in the endothelium (atherogenesis)
and the subsequent development of atherosclerosis \cite{rybicki_prediction_2009}.
In particular, low levels of shear stress, as due to disturbed flow
patterns and stagnation regions, trigger the growth of plaques. It
is therefore of great relevance to inspect the role of RBCs in modulating
the shear stress distribution on the wall of the vessel under study.

For this analysis, we employ the method described in ref. \cite{melchionna_hydrokinetic_2010}
and compute the endothelial shear stress as time averages over the
evolution of the plasma-RBC composite system. Fig. \ref{fig:ESS}
illustrates the distribution of shear stress for different hematocrit
levels. The plot reveals the strong effect of the RBCs throughout
the system and in particular the great fluctuations in one daughter
vessel. While the overall shear stress distribution is somehow preserved
at different hematocrit levels, important local modifications are
induced by RBCs. In particular, in proximity of the vessel shoulder,
the RBC structuring induces smaller values of the shear stress, followed
by larger values next to the inner side of the bifurcation. Given
these observations, it is expected that the particulate nature of
blood has significant effects in the identification and prediction
of vulnerable regions of the large-size vascular system by computational
methods. 

\begin{center}
\begin{figure}
\centering{}\includegraphics[scale=0.35]{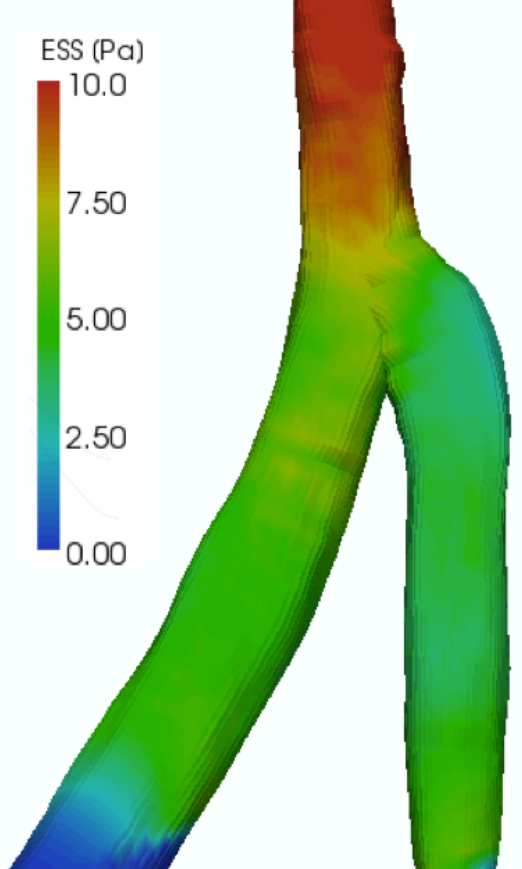}\includegraphics[scale=0.35]{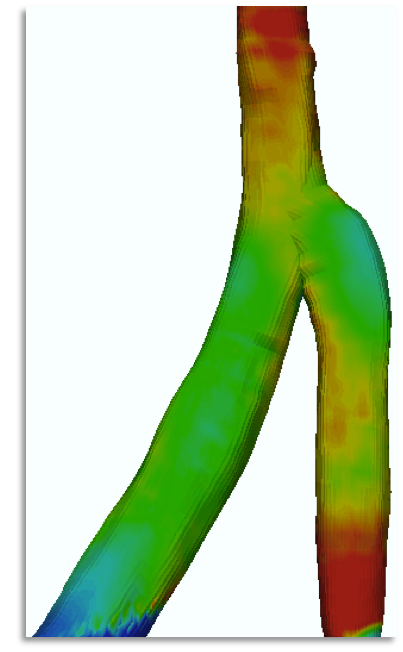}\includegraphics[scale=0.35]{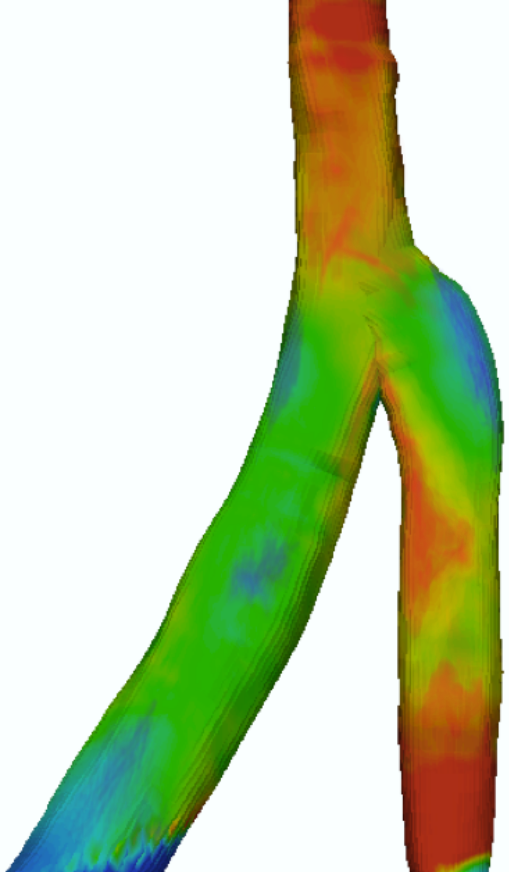}\caption{\label{fig:ESS}Time-averages of shear stress in the bifurcating channel
for pure plasma (left panel), hematocrit level of $35\,\%$ (mid panel)
and $50\,\%$ (right panel). The flow corresponds to an average shear
rate of $80\, s^{-1}$.}
\end{figure}

\par\end{center}

\section{Conclusions}

In this paper, a new model for suspended red blood cells has been
introduced with the goal of studying large-scale hemodynamics in cardiovascular
systems. The underlying idea is to represent the different responses
of the suspended bodies, emerging from the rigid-body as much as the
vesicular nature of the globule, by distinct coupling mechanisms.
These mechanisms are entirely handled at kinetic level, that is, the
dynamics of plasma and red blood cells is governed by appropriate
collisional terms that avoid to compute hydrodynamic forces and torques
via the Green's function method, as employed in Stokesian dynamics
\cite{brady_stokesian_1988}. The fundamental advantage of hydrokinetic
modeling is to avoid such expensive route and, at the same, enabling
to handle finite Reynolds conditions and complex boundaries or irregular
vessels within the simple collisional approach. At macroscopic scale,
the non-trivial rheological response emerges spontaneously as a result
of the underlying microdynamics. 

The second, crucial advantage of the proposed method is to enable
the investigation of systems of physiological relevance whose phenomenology
typically spans several decades in time and space. To this purpose,
massive parallel computers must be deployed by exploiting a numerical
method that scales over thousands of processors and above. To a large
extent, both the Lattice Boltzmann method and the red blood cell dynamics
have been proved to scale over traditional CPU-based computers such
as on Blue Gene architectures, as much as over massive assemblies
of Graphic Processing Units (see ref. \cite{peters_multiscale_2010}
and references therein).

The numerical results presented in this paper have shown that the
particulate nature of blood cannot be omitted when studying the non-trivial
rheology of the biofluid and the shear stress distribution in complex
geometries. Regions of low shear stress can appear as the hematocrit
reaches physiological levels as a result of the non-trivial organization
of red blood cells and the irregular morphology of vessels. These
results cannot can have far reaching consequences in real-life cardiovascular
applications, where the organization of red blood cells impacts both
the local flow patterns and the large-scale flow distribution in complex
vascular networks.
\begin{acknowledgments}
The author is indebted to Sauro Succi, Massimo Bernaschi, Mauro Bisson,
John Russo, Greg Lakatos, Efthimios Kaxiras, Umberto Marini Bettolo
Marconi and Howard A. Stone for enlightening discussions.
\end{acknowledgments}
\bibliographystyle{plain}
\bibliography{/users/simonemelchionna/1.TESTI/BIBTEX_DATABASE/ALLBIB}

\appendix

\section{Rigid Body Dynamics}

We report here the numerical solution of the roto-translational dynamics
for RBCs. The rotational dynamics is customarily solved in the body
frame where the equation relating angular momentum $\boldsymbol{\pi}$
and angular velocity ${\bf \Omega}$ is 
\begin{equation}
\boldsymbol{\pi}={\bf I}\boldsymbol{\Omega}
\end{equation}
where, for ease of notation, we have omitted the RBC index. The rigid-body
dynamics is given by
\begin{equation}
\dot{\boldsymbol{\Omega}}={\bf I}^{-1}\left[({\bf I}\boldsymbol{\Omega})\times\boldsymbol{\Omega}\right]+{\bf I}^{-1}\boldsymbol{\tau}
\end{equation}
and the rotational matrix obeys the equation
\begin{equation}
\dot{{\bf Q}}={\bf Q}\, skew(\boldsymbol{\Omega})
\end{equation}
where, for a vector with components ${\bf v}=(v_{n},v_{t},v_{g})$,
$skew({\bf v})$ denotes 
\begin{equation}
skew({\bf v})=\left(\begin{array}{ccc}
0 & v_{g} & -v_{t}\\
-v_{g} & 0 & v_{n}\\
v_{t} & -v_{n} & 0
\end{array}\right)
\end{equation}

In absence of hydrodynamic forces and torques, the dynamics is Hamiltonian
and the updating algorithm is given by the scheme described in ref.
\cite{dullweber_symplectic_1997}. For $\gamma_{T}=\gamma_{R}=0$
the algorithm is symplectic and provides excellent conservation of
energy. In presence of hydrodynamic forces and torques, the algorithm
is generalized as in the following steps:
\begin{enumerate}
\item Propagate linear and angular momenta for half-step according to the
combination of mechanical and hydrodynamic forces ${\bf F}^{comb}$
and torques $\boldsymbol{\tau}^{comb}$ , 
\begin{eqnarray}
{\bf \Omega}_{h/2} & = & {\bf \Omega}_{0}+\frac{h}{2}{\bf I}^{-1}\boldsymbol{\tau}_{0}^{comb}\\
{\bf V}_{h/2} & = & {\bf V}_{0}+\frac{h}{2M}{\bf F}_{0}^{comb}
\end{eqnarray}
where
\begin{eqnarray}
{\bf F}_{0}^{comb} & = & {\bf F}_{0}^{mech}+\gamma_{T}M\tilde{{\bf u}}_{0}\\
\boldsymbol{\tau}_{0}^{comb} & = & \boldsymbol{\tau}_{0}^{mech}+\gamma_{R}{\bf I}\tilde{\boldsymbol{\omega}}_{0}
\end{eqnarray}

\item Update linear and angular momenta for half-step according to the frictional
forces
\begin{eqnarray}
{\bf \Omega}_{h/2} & = & e^{-\gamma_{R}h/2}{\bf {\bf \Omega}}_{0}\\
{\bf V}_{h/2} & = & e^{-\gamma_{T}h/2}{\bf V}_{0}
\end{eqnarray}

\item Propagate positions as
\begin{equation}
{\bf R_{h}}={\bf R}_{0}+h{\bf V}_{h/2}
\end{equation}

\item Update the angular state by taking the rotation matrix ${\bf {\cal R}}(\alpha,\beta,\gamma)$,
associated to rotation around the axes $\hat{{\bf n}}$, $\hat{{\bf t}}$
and ${\bf \hat{g}}$ with angles $\alpha$, $\beta$ and $\gamma$,
respectively. Propagate the rotational state according to the sequence
of five sub-rotations

\begin{enumerate}
\item ${\bf {\cal R}}\equiv{\cal R}(\frac{h\alpha}{2},0,0)\longrightarrow\begin{cases}
\begin{array}{c}
{\bf \Omega}={\bf I}^{-1}{\bf {\cal R}}{\bf I}{\bf \Omega}\\
{\bf Q}={\bf Q}{\bf {\cal R}}^{\dagger}
\end{array}\end{cases}$
\item ${\bf {\cal R}}\equiv{\cal R}(0,\frac{h\beta}{2},0)\longrightarrow\begin{cases}
\begin{array}{c}
{\bf \Omega}={\bf I}^{-1}{\bf {\cal R}}{\bf I}{\bf \Omega}\\
{\bf Q}={\bf Q}{\bf {\cal R}}^{\dagger}
\end{array}\end{cases}$
\item ${\bf {\cal R}}\equiv{\cal R}(0,0,h\gamma)\longrightarrow\begin{cases}
\begin{array}{c}
{\bf \Omega}={\bf I}^{-1}{\bf {\cal R}}{\bf I}{\bf \Omega}\\
{\bf Q}={\bf Q}{\bf {\cal R}}^{\dagger}
\end{array}\end{cases}$
\item ${\bf {\cal R}}\equiv{\cal R}(0,\frac{h\beta}{2},0)\longrightarrow\begin{cases}
\begin{array}{c}
{\bf \Omega}={\bf I}^{-1}{\bf {\cal R}}{\bf I}{\bf \Omega}\\
{\bf Q}={\bf Q}{\bf {\cal R}}^{\dagger}
\end{array}\end{cases}$
\item ${\bf {\cal R}}\equiv{\cal R}(\frac{h\alpha}{2},0,0)\longrightarrow\begin{cases}
\begin{array}{c}
{\bf \Omega}={\bf I}^{-1}{\bf {\cal R}}{\bf I}{\bf \Omega}\\
{\bf Q}={\bf Q}{\bf {\cal R}}^{\dagger}
\end{array}\end{cases}$
\end{enumerate}
\item Compute the mechanical forces and torques from the excluded volume
interactions, ${\bf F}_{h}^{mech}$ and $\boldsymbol{\tau}_{h}^{mech}$
and the fluid velocity and vorticity via Eqs.(\ref{eq:vel}-\ref{eq:vort}),
and construct the combinations 
\begin{eqnarray}
{\bf F}_{h}^{comb} & = & {\bf F}_{h}^{mech}+\gamma_{T}M\tilde{{\bf u}}_{h}\\
\boldsymbol{\tau}_{h}^{comb} & = & \boldsymbol{\tau}_{h}^{mech}+\gamma_{R}{\bf I}\tilde{\boldsymbol{\omega}}_{h}
\end{eqnarray}

\item Update linear and angular momenta for half-step according to the frictional
forces
\begin{eqnarray}
{\bf \Omega}_{h} & = & e^{-\gamma_{R}h/2}{\bf {\bf \Omega}}_{h/2}\\
{\bf V}_{h} & = & e^{-\gamma_{T}h/2}{\bf V}_{h/2}
\end{eqnarray}

\item Finalize the linear and angular momenta for half-step according to
combined forces and torques
\begin{eqnarray}
{\bf V}_{h} & = & {\bf V}_{h/2}+\frac{h}{2M}{\bf F}_{h}^{comb}\\
{\bf \Omega}_{h} & = & {\bf \Omega}_{h/2}+\frac{h}{2}{\bf I}^{-1}\boldsymbol{\tau}_{h}^{comb}
\end{eqnarray}

\end{enumerate}

\section{Gay-Berne forces and torques}

The calculation of mechanical forces and torques according to the
Gay-Berne potential, eq. (\ref{eq:gayberne}), is reported here. For
the interacting pair $i,j$ , the pairwise mechanical force is given
by
\begin{equation}
{\bf {\bf F}}_{ij}^{GB}=-\frac{\partial u_{ij}}{\partial R_{ij}}\hat{{\bf R}}_{ij}-\frac{1}{R_{ij}}\frac{\partial u_{ij}}{\partial\phi_{ij}}\frac{\partial\phi_{ij}}{\partial\hat{{\bf R}}_{ij}}\cdot\left({\bf 1}-\hat{{\bf R}}_{ij}\hat{{\bf R}}_{ij}\right)=-\frac{\partial u_{ij}}{\partial R_{ij}}\hat{{\bf R}}_{ij}-\frac{1}{R_{ij}^{2}}\frac{\partial u_{ij}}{\partial\phi_{ij}}\left[\boldsymbol{\kappa}_{ij}-(\boldsymbol{\kappa}_{ij}\cdot\hat{{\bf R}}_{ij})\hat{{\bf R}}_{ij}\right]\label{eq:forcegb}
\end{equation}
where the following vector has been defined
\begin{equation}
\boldsymbol{\kappa}_{ij}={\bf H}_{ij}^{-1}\cdot{\bf R}_{ij}
\end{equation}
The terms appearing in eq. (\ref{eq:forcegb}) are computed as
\begin{equation}
\frac{\partial u_{ij}}{\partial R_{ij}}=-24\epsilon_{0}\left[\frac{\left(2\rho_{ij}^{-13}-\rho_{ij}^{-7}\right)}{\sigma_{ij}^{min}}\right]
\end{equation}
and
\begin{equation}
\frac{\partial u_{ij}}{\partial\phi_{ij}}=-24\epsilon_{0}\left[\frac{\left(2\rho_{ij}^{-13}-\rho_{ij}^{-7}\right)\sigma_{ij}^{3}}{2\sigma_{ij}^{min}}\right]
\end{equation}

The torques are calculated according to the prescription in ref. \cite{allen_expressions_2006}
and transformed back to the body frame 
\begin{equation}
\boldsymbol{\tau}_{i}^{GB}={\bf Q}_{i}^{T}({\bf R}_{ci}\times{\bf f}_{\kappa_{ij}})
\end{equation}
and 
\begin{equation}
\boldsymbol{\tau}_{j}^{GB}={\bf Q}_{j}^{T}({\bf R}_{cj}\times{\bf f}_{\kappa_{ij}})
\end{equation}
where
\begin{equation}
{\bf f}_{\kappa_{ij}}=-\frac{1}{R_{ij}}\frac{\partial u_{ij}}{\partial\phi_{ij}}\boldsymbol{\kappa}_{ij}
\end{equation}
and 
\begin{eqnarray*}
{\bf R}_{ci} & = & {\bf R}_{c}-{\bf R}_{i}\\
{\bf R}_{cj} & = & {\bf R}_{c}-{\bf R}_{j}\\
{\bf R}_{c} & = & {\bf R}_{i}-{\bf A}_{i}\boldsymbol{\kappa}_{ij}
\end{eqnarray*}
It is easy to show that torques obey conservation of angular momentum
since $\boldsymbol{\tau}_{i}^{GB}+\boldsymbol{\tau}_{j}^{GB}=-{\bf R}_{ij}\times{\bf F}_{ij}^{GB}$.
\end{document}